\documentclass[prl,superscriptaddress,twocolumn,floatfix]{revtex4}
\usepackage[latin1]{inputenc}  
\usepackage[austrian,english,french]{babel}
\usepackage{graphicx}
\usepackage{amsmath}
\usepackage{amssymb}
\usepackage{amscd}
\usepackage[sort&compress]{natbib}

\begin{document}
\selectlanguage{english}

\title{A Novel Protocol-Authentication Algorithm Ruling Out a
Man-in-the-Middle Attack in Quantum Cryptography}
\author{M. Peev}
\email{momchil.peev@arcs.ac.at}
\author{M. N\"olle}
\author{O. Maurhardt}
\author{T. Lor\"unser}
\author{M. Suda}
\affiliation{ARC Seibersdorf Research GmbH, A-1220 Tech Gate
Vienna, Donau-City-Stra{\ss}e 1 / 4. OG, Austria}

\author{A. Poppe}
\author{R. Ursin}
\author{A. Fedrizzi}
\affiliation{Institut für Experimentalphysik, Universität Wien,
 Austria}

\author{A. Zeilinger}
\affiliation{Institut für Experimentalphysik, Universität Wien,
 Austria}
\affiliation{Institute for Quantum Optics and Quantum Information,
Austrian Academy of Sciences}
\date{\today}

\begin{abstract}
In this work we review the security vulnerability of Quantum
Cryptography with respect to  "man-in-the-middle attacks" and the
standard authentication methods applied to counteract these
attacks. We further propose a modified authentication algorithm
which features higher efficiency with respect to consumption of
mutual secret bits.
 \vspace{0cm}
\end{abstract}
 \maketitle
\section{Introduction}
Quantum Key Distribution (QKD) or "quantum cryptography" is a
Quantum Mechanics based cryptographic primitive which, in
principle, holds the potential of absolutely secure communication
that cannot be compromised by any eavesdropping technique. The
strength of the QKD primitive is the unconditionally secure
simultaneous generation of two identical bit streams at two
distinct locations which subsequently could be used as a key in
symmetric (unconditionally or computationally secure) encryption
schemes. However, it is well known that QKD requires a public
channel with trusted integrity as otherwise a potential adversary
(Eve) can easily mount a "man-in-the-middle attack". In case the
eavesdropper can manipulate messages on the public channel there
is no way to guarantee that in the course of a QKD protocol the
two legitimate communication parties (Alice and Bob) are really
exchanging the messages they are sending to each other. Eve can
simply cut the quantum channel and subsequently communicate over
both the quantum and the public channels with Bob as if she would
be Alice and with Alice as if she would be Bob. Eventually, she
would thus share two independent keys with the two legitimate
parties and gain full control of all the subsequently transmitted
encrypted information without being noticed at all. The described
type of attack can be counteracted by authenticating the QKD
protocol messages transmitted over the public channel. Basically
public key authentication methods and symmetric key authentication
methods can be used (see Ref.~\cite{Pat} for a  discussion of the
relative merits and drawbacks of these methods). It is however
straightforward to notice that unconditionally secure key
generation by means of QKD is only feasible if it is combined with
methods providing unconditionally secure authentication. Standard
public key methods are automatically ruled out if one would stick
to this requirement as the latter are only computationally secure
and potentially subject to cryptanalysis by means of quantum
computers. Therefore, already in Ref.~\cite{Ben} it was proposed
to use unconditionally secure symmetric message authentication
methods as e.g. developed in Ref.~\cite{Weg} to ensure the
integrity of the public channel. The main idea of the application
of these methods in QKD is to intertwine the transcript of the
public channel communication with an independent secret, which the
two legitimate parties share and on this basis provide a mechanism
for authenticating this communication. Alice and Bob need
therefore an initial secret key, which they use only once.
Subsequently in each QKD session they repeatedly renew the mutual
secret by reserving part of the newly generated key. This key is
to be used for channel authentication purposes in the next
session. This paradigm has been elaborated in subsequent
publications\cite{Dus,Gilb}. It should be noted that while thus
the unconditional security of QKD is retained, it is basically
degraded from a secret key generation scheme in the strict sense
to a secret key growing technique.

In what follows we restrict our discussion to symmetric key
message authentication methods and, similar to Wegman and
Carter\cite{Weg}, base our approach on strongly universal$_{2}$
functions. In Section 2 we discuss a general method for producing
message authentication tags using only a moderate amount of the
secret key. In Section 3 we briefly discuss the details of the
authentication algorithm in relation to the QKD protocol. We also
present a modular integrated software library implementing full
scale QKD-protocols including public channel authentication used
in the framework of a recent quantum cryptographic
experiment\cite{Pop}.

\section{Message Authentication Primitive}
A broad class of unconditionally secure symmetric key message
authentication approaches follow the method described in
Ref.~\cite{Weg}. This method is independent of the context in
which authentication is applied and therefore we refer to it in
what follows as the authentication primitive. Before discussing
the primitive itself we shortly review the foundations of message
authentication by means of a family of strongly universal$_{2}$
functions. Let $H_{A}$ be such a family of functions which maps
the set of all messages $A$, typically the set of binary strings
of length $m$, to the set of all authentication tags $B$,
typically the set of binary strings of length $n<m$. One can then
authenticate a message by sending an authentication tag in
addition to the message itself over the communication channel. An
adversary willing to manipulate the original message must also be
able to produce the proper tag for the manipulated message. The
authentication system is unbreakable with probability $p$ when $B$
is chosen to have at least $1/p$ elements\cite{Weg}. The term
"unbreakable with probability $p$" is used in the following sense:
If a message from $l{\in}A$ yields a tag $t{\in}B$ through a
randomly chosen function $f{\in}H_{A}$, $t=f(l)$, and if an
eavesdropper knows $m$ and $t$ but not $f$, she has only a
probability lower than $p$ to find the proper tag $t'$ of a
different message $l'$, with respect to $f$, $t'=f(l')$. The
legitimate parties share a secret key, which is used as an index
in the function space $H_{A}$. In this sense the secret sharing is
symmetric. The secret can be used only once. The problem with this
basic approach is that most of the well known families of strongly
universal$_{2}$ functions are typically larger than the space of
all messages. Therefore, the key needed to authenticate a message
is longer than the message itself. This is a particular problem in
quantum cryptography, where the key growth factor directly depends
on the portion of generated key, reserved for a subsequent
authentication. While it is necessary to minimize the length of
the messages to be authenticated as discussed in Section 3, it is
also strongly desirable to restrict the space of applied hash
functions, reduce the secure key consumption for authentication
purposes and thus get efficient authentication methods. At the
expense of increasing the "security parameter" $p$ to $2p$, Wegman
and Carter propose a method for building a relatively restricted
family of almost strongly universal$_{2}$ hash
functions\cite{Weg}, which uses a basic class of strongly
universal$_{2}$ hash functions into intermediate spaces as a
kernel.  Wegman and Carter choose a specific multiplicative family
of hash functions (denoted as $H_{1}$ in Ref.~\cite{Car}), to map
strings of length $2s$ to those of length $s$, where

\begin{equation}
s = n + {\log}_{2}{\log}_{2}m\,\,\,.
\end{equation}

Note that by definition the cardinality of this class, being a
function of s, only slowly grows with $m$. The original message
$l$ is then divided into substrings of a defined length $2s$ and a
randomly chosen hash function from the mentioned class is applied
to the substrings. The set of resulting tags is then concatenated
to produce an intermediate message. The latter is then once again
subdivided into substrings of the length $2s$ and a new hash
function from the described family is applied to each string. This
process is applied until only one tag remains. The lower order $n$
bits are taken for the final authentication tag $t$. One can
show\cite{Weg} that this method defines an almost strongly
universal$_{2}$ family of functions from $A$ onto $B$. Wegman and
Carter also prove that the key length needed to index this family
is

\begin{equation}
k = 4s{\log}_{2}m\,\,\,.
\end{equation}

This method constitutes a general primitive for symmetric key
authentication. The definition $f$ of the almost strongly
universal$_{2}$ class of hash functions is independent of the
underlying kernel class of intermediate strongly universal$_{2}$
functions and any such class can be used. The authentication of
the public channel in QKD discussed so far in literature (see e.g.
Refs.~\cite{Ben}, ~\cite{Gilb} and ~\cite{Bou}) are almost
exclusively based on the discussed primitive developed in
Ref.~\cite{Weg}, including the choice of the basic intermediate
class of strongly universal$_{2}$ ($2s$ to $s$) hash functions. It
is obvious that this method is suitable for authenticating long
messages. As an example, for authentication tags which are 64 bits
long the message length exceeds the key length if the former is
longer than 3138 Bits. For messages longer than 20000 Bits the
message length exceeds the key lengths already by a factor of
four. However, in certain settings, and in particular in the QKD
case, it is highly relevant to have an efficient authentication
primitive also for short messages. To this end we propose a new
primitive, which includes a two step procedure. First of all one
maps the initial message $l$ from $A$ to $Z$, where $Z$ is the set
of all binary strings of length $r$ ($m>r>n$), by means of a
single publicly known hash function $f_{0}$ so that $z =
f_{0}(l)$. The second step is a direct application of the basic
approach as discussed above. One sends $m$ over the communication
channel alongside with $t = f(z)$, where $f$ is a randomly chosen
secret strongly universal$_{2}$ hash function from $H_{Z}$ mapping
$Z$ onto $B$. We discuss first the security of this primitive and
then assess the amount of secret key needed for its
implementation. The security of the primitive is given by the
probability $p$ of an adversary to produce a proper authentication
tag for a modified message (cf. the discussion above). Obviously

\begin{equation}
p = p_{1} + p_{2}\,,\,\,\,\,p_{2}=1/|B|\,\,\,.
\end{equation}

Here, $p_{2}$ is the probability for the eavesdropper to break the
strongly universal$_{2}$ family $H_{Z}$ (see Ref.~\cite{Weg})
while $p_{1}$ is the probability that the initial message and the
modified message yield the same tag $q$ under the chosen fixed
hash function $f_{0}$:

\begin{equation}
p_{1} = max_{l}\Big( Pr{\{}f_{0}(l) = f_{0}(l') | l \neq
l'{\}}\Big)\,\,\,.
\end{equation}

Clearly all messages $A_{0} = f_{0}^{-1}(z)$ yield the same
authentication tag $z$ and thus

\begin{equation}
p_{1} = max_{l}\Big(Pr \left\{l'{\in}\bar{A}_{0}%
=\{A_{0}{\setminus}\,l\}\right\}\Big)\,\,\,.
\end{equation}

In case $A_{0}$ is independent of the choice of $l$ and all $l$
are equally probable (the distribution of meaningful messages in
the space of all bit strings is uniform) then

\begin{eqnarray}
\nonumber & p_{1} = (|A|/|Z|-1)/|A|<1/|Z|\,\,\, \mbox{for
all values of }\,|A|, \\
& p<1/|B|+ 1/|Z|\,.
\end{eqnarray}

In addition to the two basic assumptions in the derivation of
expression Eq.(6) one should note that we implicitly assume that
the message $l'$ is random and fixed, i.e. the eavesdropper can
not chose  the manipulation message at will. While this can not be
taken for granted in general, in the case of man-in-the-middle
attacks in quantum cryptography $l$ an $l'$ are definitely
randomly and independently fixed beside the scope of influence of
the adversary. (In this case of a man-in-the middle attack $l$ and
$l'$ are protocol extracts from the communication between Alice
and Eve and Eve and Bob respectively, whereby these are generated
through physical random processes and Eve has no opportunity at
all to change either $l$ or $l'$.) The assumption that $A_{0}$ is
independent from $l$ can be guaranteed by any suitably chosen\\
hash function that constitutes a homomorphism of $A$ onto $Z$.
Finally the assumption of an uniform distribution of all possible
messages depends on the choice of the protocol extracts and can
not  always be granted. However, one can initially perform a
uniform randomizing operation e.g. by means of XORing the message
$l$ with a completely random bit string of the same dimension. The
latter can be obtained by means of deterministic pseudorandom
generator whereby a number of secret bits from the joint secret
are used for the seed. The application of other appropriate
uniform randomizers, possibly integrated in the definition of
$f_{0}$, is also feasible.

We would now point out that the secret key needed in this approach
is exactly the number of bits needed to index the family $H_{Z}$.
Obviously if $r$ (the dimension   of $Z$) is chosen to be moderate
and an appropriate restricted strongly universal$_{2}$ class is
selected then the amount of secret key required can also be
reduced. To estimate  this amount exactly one needs to specify the
function family applied. We choose the set of affine
transformations:

\begin{eqnarray}
\nonumber  & H_{Z}(\cdot)={\{}f=(\Phi,\beta)|\Phi- \mbox{ all}\,(r{\times}m)\, \mbox{binary} \,\\
\nonumber & \mbox{\hspace{0pt} Toeplitz matrices;} \, \beta - \mbox{all}\,(m{\times}1)\,\mbox{binary vectors} {\}}\, ,\,\,\\
& f(z)={\Phi}z + \beta\,\,\,mod(2)\,\,.
\end{eqnarray}

This function family\cite{Kraw,Man} is strongly universal$_{2}$ as
shown in Ref.~\cite{Man} and is indexed by $r+2\times n-1$
parameters. For $r$=256 and $n$=64, obviously the message length
exceeds the required secret key already for strings longer than
384 bits. In contrast to the primitive used by Wegman and Carter
this amount is constant by definition and does not increase with
$m$.

\section{Protocol Implementation}
\vspace*{-1pt} In Section 2 above we have only   given a general
description of a new primitive suitable for authentication in QKD
settings, leaving
 the question of the exact protocol extracts to be
authenticated completely aside. It is beyond the scope of the
present paper to address this topic in detail. This issue has been
however thoroughly discussed in Refs.~\cite{Dus} and ~\cite{Gilb}.
Certainly an authentication of the  full protocol transcript is
one (inefficient) extreme possibility. In Ref.~\cite{Gilb} it is
shown that authenticating the sifting phase discussion and the
results of the error correction phase is sufficient. In this
reference it is also suggested that (the relevant parts) of the
transcript are not authenticated at once but rather the bit
strings to be authenticated are separately processed as soon as
they are generated in the respective QKD protocol phases. A
particular advantage of this approach with respect to the
cryptographic primitive proposed in Section 2 above, is that all
the bit strings to be authenticated are randomly and uniformly
distributed in the space of all possible strings of corresponding
dimension. Thus, if this protocol is employed, an initial
randomization is not required for a secure application the
primitive described above.

 We have implemented an authentication algorithm
based on the primitive presented in this paper, whereby,
provisionally, SHA is used as the initial hash function. This
algorithm is a part of a constantly developed modular software-set
up, which is integrated in the framework of an embedded general
purpose QKD hardware-software prototype dedicated to data
acquisition and subsequent QKD-protocol processing and data
encryption (currently AES and One Time Pad are implemented). A
public demonstration of the functionality of this QKD prototype
together with an optic segment implementing entangled-photon  key
generation took recently place in the form of a "Q-Banking"
transaction, which was carried out between two buildings in
Vienna, Austria - the Rathaus (city hall) and the seat of Bank
Austria Creditanstalt\cite{Pop}. The current version of the
software set-up is designed in the form of a C library "QKD III"
which allows application by choice of alternative
quantum-acquisition protocols, error correction, privacy
amplification and authentication algorithms and can alternatively
be compiled for usage in  PC or embedded environments. The QKD
protocol implemented, in contrast to an earlier version used in
the "Q-Banking" experiment, follows the approach suggested in
Ref.~\cite{Gilb}. This protocol is prone by design against a
potential loophole in this earlier version. The latter is
discussed in detail in Ref.~\cite{Muel}. \nopagebreak

\section*{Acknowledgements}

We would thank J\"orn M\"uller-Quade and Rainer Stein- \newpage
\noindent wandt for helpful discussions and comments. This work
has been supported by the FIT/IT project PRODEQUAC (Nr. 806015 -
Austrian Ministry of Transportation and Innovation) and
 the EC/IST integrated project SECOQC (Contract No 506813).

\end{document}